\documentclass[aps,prl,twocolumn,showpacs,superscriptaddress,nofootinbib]{revtex4-1}
\usepackage{epsfig}
\usepackage{color}
\usepackage{endnotes}
\usepackage{natbib}
\usepackage[colorlinks,breaklinks]{hyperref}
\usepackage{nicefrac}
\usepackage{capt-of}
\usepackage{bm}% bold math
\usepackage{dcolumn}
\usepackage{graphicx}
\usepackage{graphics}% Include figure files
\usepackage{amsmath}
\usepackage{amssymb}
\usepackage{xcolor}
\usepackage{relsize}
\newcommand{\bs}[1]{\ensuremath{\boldsymbol{#1}}}
\newcommand{\be}{\begin{equation}}
\newcommand{\ee}{\end{equation}}
\newcommand{\bea}{\begin{eqnarray}}
\newcommand{\eea}{\end{eqnarray}}

\newcommand{\sgmvec}{\ensuremath{\boldsymbol{\sigma}}}
\newcommand{\tauvec}{\ensuremath{\boldsymbol{\tau}}}
\newcommand{\nopi}{\pi\hspace{-6pt}/}

\newcommand{\simge}{\hspace*{0.2em}\raisebox{0.5ex}{$>$}
     \hspace{-1em}\raisebox{-0.3em}{$\sim$}\hspace*{0.2em}}

\begin{document}

\title{Effective Field Theory for Lattice Nuclei}

\author{N.~Barnea}
%\email{nir@phys.huji.ac.il}
\affiliation{Racah Institute of Physics, The Hebrew University, Jerusalem 91904, Israel}

\author{L.~Contessi}
%\email{}
\affiliation{Physics Department, University of Trento, via Sommarive 14, I-38123 Trento, Italy}

\author{D.~Gazit}
%\email{doron.gazit@phys.huji.ac.il}
\affiliation{Racah Institute of Physics, The Hebrew University, Jerusalem 91904, Israel}

\author{F.~Pederiva}
%\email{}
\affiliation{Physics Department, University of Trento, via Sommarive 14, I-38123 Trento, Italy}
\affiliation{INFN-TIFPA Trento Institute for Fundamental Physics and Applications, Trento, Italy}

\author{U.~van~Kolck}
%\email{vankolck@ipno.in2p3.fr, vankolck@physics.arizona.edu}
\affiliation{Institut de Physique Nucl\'eaire, CNRS/IN2P3, 
Universit\'e Paris-Sud, F-91406 Orsay, France}
\affiliation{Department of Physics, University of Arizona,
Tucson, AZ 85721, USA}

\date{\today}

%===========================================================
\begin{abstract}
%Stimulated by recent lattice quantum chromo dynamics (LQCD) simulations of few
%baryon systems including 
%the deutron, triton and the $\alpha$-particle  
%we device an effective field theory (EFT) to analyze these lattice nuclei.
%Confined at the moment to pion mass much heavier then the physical pion mass,
%we argue that pion-less EFT is the appropriate theory for these systems.
%We apply the theory to study available LQCD measurments at $m_{\pi}\approx
%800$ MeV, and to predict mass 5 and 6 nuclear binding energies.
We show how nuclear effective field theory (EFT) 
and {\it ab initio} nuclear-structure methods
can turn input from lattice quantum chromodynamics (LQCD) 
into predictions for the properties of nuclei.
We argue that pionless EFT is the appropriate theory 
to describe the light nuclei obtained in recent LQCD simulations 
carried out at 
pion masses much heavier than the physical pion mass.
We solve the EFT using
the effective-interaction hyperspherical harmonics
and auxiliary-field diffusion Monte Carlo methods.
Fitting the three leading-order EFT parameters to the deuteron,
dineutron and triton LQCD energies at $m_{\pi}\approx 800$ MeV, 
we reproduce the corresponding alpha-particle binding and predict the 
binding energies of mass-5 and 6 ground states.
\end{abstract}


\maketitle

%===========================================================
% Introduction
%===========================================================
{\it Introduction} --
Understanding the low-energy dynamics of quantum chromodynamics
(QCD), which underlies the structure of nuclei, is a longstanding 
challenge posed by its non-perturbative nature. 
After many years of development, 
lattice QCD (LQCD) simulations are fulfilling their 
promise of calculating static and dynamical
quantities with controlled approximations.
Progress has reached the point where meson and 
single-baryon properties can be predicted quite 
accurately, see for example Ref. \cite{Alvarez13}.
Following the pioneering studies in quenched \cite{LQCDpioneers}
and fully-dynamical \cite{firstNPLQCD}  LQCD,
a substantial effort is now in progress to study light
nuclei 
%and to extract information on the nuclear force
\cite{lqcd_nn_force,Yamazaki12,NPLQCD13a,NPLQCD13b}. 
%This effort is highly challenging as 
Multinucleon systems are significantly
more difficult to calculate than single-baryon states,
as they are more complex, demand larger lattice volumes, and
better accuracy to account for the fine-tuning of the nuclear force.  
At heavier light-quark masses, 
the formation of quark-antiquark pairs is suppressed,
the computational resources required to generate LQCD configurations 
are reduced,
and the signal-to-noise ratio in multinucleon correlation
function improves \cite{NPLQCD13b}.  
Therefore, present multinucleon LQCD simulations are performed
at heavy up and down
quark masses, which result in
unphysical values for hadronic quantities.
%masses of the pion and nucleons.
Once lattice artifacts
%volume effects 
are accounted for 
using large enough volumes and extrapolating to the continuum, 
LQCD results depend on a single parameter, the pion mass $m_{\pi}$.
However, sufficiently large volumes are harder to achieve
as the number of nucleons increases due to the 
the saturation of nuclear forces.

A hadronic effective field theory (EFT) that incorporates chiral
symmetry (chiral EFT)
provides a tool to extrapolate LQCD results 
%for multinucleon systems 
to a smaller, more realistic 
%quark masses
pion mass \cite{BBSvK,firstNPLQCD}. 
Here we show how EFTs,
combined 
with {\it ab initio} methods for
the solution of the Schr\"odinger equation,
provide 
%an alternative 
a way to extend LQCD results also to the larger distances involved
in nuclei with 
%a large number of 
several nucleons. Of course, solving 
the nuclear many-body problem is not a small challenge, yet it is considerably
simpler than solving QCD on the lattice. 
 
We devise an EFT for 
existing lattice nuclei, that is, nuclei 
composed of neutrons and protons living in a world where $m_{\pi}$ is much 
larger than the physical pion mass. 
%For such masses, 
Pion effects can be considered short-ranged,
and the appropriate theory is pionless EFT ($\nopi$EFT),
an EFT based on the most general dynamics among nucleons
which is consistent with the symmetries of QCD.
(For a review, see {\it e.g.} Ref. \cite{paulo}).
We solve this EFT in leading order (LO) using the
effective-interaction hyperspherical harmonics (EIHH) method \cite{EIHH}
for systems with $A\le 6$ nucleons,
and 
the auxiliary-field diffusion Monte Carlo (AFDMC) method \cite{AFDMC1,AFDMC2}
for $A\ge 4$.
Binding energies of nuclei with $A\le 3$ are used as input.
The energy of the $A=4$ ground state provides a consistency check between
both {\it ab initio} methods, and between them and LQCD.
Binding energies for $A\ge 5$ are predictions that extend 
LQCD into new territory.
In order to evaluate the feasibility of our approach,
we present here the first
analysis of the problem using recent multinucleon LQCD results at
$m_{\pi}=805\; \rm{MeV}$ from the NPLQCD collaboration \cite{NPLQCD13a}.
Table \ref{tbl:Data} summarizes nucleon and light nuclear data in nature
and in the LQCD world, including our results.
%Our results are summarized in Table \ref{tbl:Data}.

\begin{table}[bt]
\begin{center}
\caption{Available experimental and LQCD data
at various values of the pion mass [MeV], and our results:
%lattice nuclei:
the neutron and proton masses and binding
energies of the lightest nuclei [MeV]. 
Fitted values are marked with *.
Error estimates are discussed in the text.}
\label{tbl:Data}
\begin{tabular}{ccccc}
%{c@{\hspace{5mm}} c@{\hspace{5mm}} c@{\hspace{5mm}} c@{\hspace{5mm} c@{\hspace{5mm}} }
\hline\hline
$m_{\pi}$     &  $140$      & $510$ & $805$ & $805$ \\
Nucleus & [Nature] &  \cite{Yamazaki12}     & \cite{NPLQCD13a} 
& [This work]\\
\hline
n      &  939.6 &  1320.0     & 1634.0  & 1634.0\\ 
p      &  938.3 &  1320.0     & 1634.0  & 1634.0\\
\hline                  
nn     &   -    &    7.4 $\pm$ 1.4  &   15.9 $\pm$ 3.8   & 15.9 $\pm$ 3.8 * \\
D      &  2.224 &   11.5 $\pm$ 1.3  &   19.5 $\pm$ 4.8   & 19.5 $\pm$ 4.8 *\\
$^3$n  &   -    &                   &                    &   -         \\
$^3$H  &  8.482 &   20.3 $\pm$ 4.5  &   53.9 $\pm$ 10.7  &  53.9 $\pm$ 10.7 *\\
$^3$He &  7.718 &   20.3 $\pm$ 4.5  &   53.9 $\pm$ 10.7  &  53.9 $\pm$ 10.7\\
$^4$He &  28.30 &   43.0 $\pm$ 14.4 &  107.0 $\pm$ 24.2  &  89 $\pm$ 36\\
$^5$He &  27.50 &                   &                    &  98 $\pm$ 39\\
$^5$Li &  26.61 &                   &                    &  98 $\pm$ 39\\
%$^6$He &  29.12 &                   &                    &  \\
$^6$Li &  32.00 &                   &                    & 122 $\pm$ 50\\
\hline\hline
\end{tabular}
\end{center}
\end{table}

%{\bf Experimental and our values in Table \ref{tbl:Data} need checking.}

The modern approach to nuclear physics deploys 
{\it ab initio} methods
such as EIHH and ADFMC in the solution of
%This is practically equivalent to the modern approach to nuclear physics, 
%in which one uses 
chiral EFT 
%to construct a Lagrangian which is composed of 
%all operators consistent with the symmetries of the fundamental theory, QCD, 
with coupling constants tuned to experimental few-body data.
%fit observables. 
Since the latter
%experimental measurements of nuclear properties 
are replaced here by LQCD data,
%in order to determine the EFT parameters,
our approach illustrates how eventually one will be able
to derive the structure of real nuclei directly from QCD. 
%Our results also show that EFT
%is appropriate to describe the nuclear regime.
Our method can be extended straightforwardly to 
hypernuclei.

%===========================================================
% EFT - energy scales
%===========================================================

{\it Effective Field Theory} --
Identification of the relevant energy scales and the selection of the
appropriate degrees of freedom is 
essential
% fundamental 
for a successful application 
of EFT to a physical problem. 
In Table~\ref{tbl:EnergyScales} we present the 
relevant energy scales for
natural nuclear physics and for lattice nuclei, 
%with
%$m_{\pi}=400\;\rm{MeV}$ 
%$m_{\pi}\sim 500, 800 \;\rm{MeV}$,
%and $\sim 800\;\rm{MeV}$.
as inferred from Table \ref{tbl:Data}. 

\begin{table}[bt]
\begin{center} 
\caption{Variation of the nuclear energy scales with the pion mass.}
%LQCD data from Refs. \cite{Alvarez13,NPLQCD13a}.}
\label{tbl:EnergyScales}
\begin{tabular}{cccc}
%{c@{\hspace{5mm}} c@{\hspace{5mm}} c@{\hspace{5mm}} c@{\hspace{5mm}}} 
\hline \hline
Scale & { $m_{\pi}\sim 140$ MeV} & { $m_{\pi}\sim 500$ MeV} & { $m_{\pi}\sim 800$ MeV} \\
\hline
 $M_{QCD}$    & 1000 MeV    & 1300 MeV  & 1600 MeV   \\
 $M_{\Delta}$  & 300 MeV     &  300 MeV  &  180 MeV   \\
 $M_{\pi}$    & 140 MeV     &  500 MeV  &  800 MeV   \\
 $M_{ope}$    & 20 MeV      &  200 MeV  &  400 MeV   \\
 $M_{nuc}$    & 10 MeV      &   15 MeV   &  25 MeV   \\
%10-25 Mev &   25 MeV   \\
\hline \hline
\end{tabular}
\end{center}
\end{table}

In nature,
nuclear physics comprises several scales.
%consists several relevant mass scales. 
The higher is the QCD scale 
$M_{QCD}\sim m_N \sim 1\;\rm{GeV}$ that characterizes the nucleon
mass $m_N$, most other hadron masses, and the chiral-symmetry-breaking scale.
%are also of the same order of magnitude. 
The second and third scales are given by the energies of the lightest 
nucleon excitation and meson, respectively,
%the lowest nucleon excitation energy,
%associated with the Delta-nucleon mass difference, 
$M_{\Delta}\sim m_{\Delta}-m_N\sim 300\;\rm{MeV}$
associated with the Delta-nucleon mass difference and
%A third scale is the lightest physical meson mass,
$M_{\pi}\sim m_{\pi}\sim 140\;\rm{MeV}$.
%associated with the physical pion mass.
Both these scales are numerically 
%this scale is
not very different from
%the magnitude of 
the pion decay constant and the Fermi
momentum in heavy nuclei.
Another energy scale, which we call the one-pion-exchange scale, 
emerges when the inverse pion Compton wavelength
%pion momentum $q_{\pi}=m_{\pi}/\hbar c$ 
is combined with the QCD mass scale,
$M_{ope}\sim m_{\pi}^2/m_N \sim 20\;\rm{MeV}$.
%$\Lambda_{ope}\sim q_{\pi}^2/M_n \sim 20\;\rm{MeV}$. 
This is also the characteristic magnitude of the nuclear binding
energy per nucleon, $M_{nuc}\sim B/A$.
%thus $\Lambda_{ope}\sim\Lambda_{nuc}\sim B/A$.

For lattice nuclei these scales can be different.
%vary with the pion mass $m_{\pi}$. 
We observe that 
%while both nucleon and Delta masses
%increase \cite{Alvarez13}, Delta effects become relatively
%more important.
%The increase in nucleon mass is not as pronounced
%as in $m_\pi$ itself, and $M_{ope}$ grows quickly,
%much faster than $M_{nuc}$.
%LQCD measurements \cite{NPLQCD13a} of the $^4$He
%$\alpha$ particle 
%binding energy at
%$m_{\pi}=805\;\rm{MeV}$ yield $B/A \sim 25\;\rm{MeV}$,
% Comparing this result
%with the physical $B/A$ we conclude 
%so that, 
%at least for light nuclei,
%the effect of $m_{\pi}$ on $M_{nuc}$ is rather moderate. 
%Comparing LQCD nucleon mass measurments for different pion masses 
%\cite{Alvarez13} we can conclude that 
%$\Lambda_{ope} \sim m_{\pi}^2/M_n(m_{\pi})\sim m_{\pi}^2/M_n(0) $
%for pion masses below $1\;\rm{GeV}$. 
%Consequently, for increasing $m_{\pi}$ 
%the one-pion-exchange scale 
%$M_{ope}$ 
%grows much faster than the nuclear scale,
%$M_{nuc}$, and
the approximate degeneracy between $M_{ope}$ and $M_{nuc}$,
so important in nature, is removed and
%at large pion masses.
a clear separation develops
%of scales 
between $M_{nuc}$ and the other scales.
%is developed for $m_{\pi}\geq 300\;\rm{MeV}$.
%Consequently, the
Barring a dramatic, unforeseen relative decrease in the mass of another nucleon
excitation or meson,
nucleons are expected to be the only relevant degrees of
freedom for low-energy lattice nuclei.
An EFT involving 
the most general dynamics 
of only the non-relativistic four-component nucleon field
(two spin and two isospin states),
%$N=(p\!\uparrow, p\!\downarrow, n\!\uparrow, n\!\downarrow)$,
%$N=(p\uparrow, p\downarrow, n\uparrow, n\downarrow)$,
%where $p$ ($n$) denotes a proton (neutron)
%and $\uparrow$ ($\downarrow$) a spin up (down),
%of only nucleons, 
without any mesons,
is the appropriate theory for these systems.
%{\it The Lagrangian} -- 
%Written in terms of the non-relativistic four-component nucleon field
%$N=(p\!\uparrow, p\!\downarrow, n\!\uparrow, n\!\downarrow)$,
%$N=(p\uparrow, p\downarrow, n\uparrow, n\downarrow)$,
%where $p$ ($n$) denotes a proton (neutron)
%and $\uparrow$ ($\downarrow$) a spin up (down),
For a process with external momenta $Q\sim \sqrt{m_N M_{nuc}}$
it produces the same $S$ matrix as QCD, but
in an expansion in $Q/M$, where $M$ is the
typical scale of higher-energy effects.
%The effective Lagrangian $\Lag$ is an expansion in $(Q/\Lambda)$ 
%where $\Lambda$ is the
%typical scale of higher energy effects \cite{Bira99} and 
%$Q\sim\Lambda_{nuc}$
%is the typical nuclear scale. 
%Written in terms of the non realtivistic four component nucleon field
%$N=(p\!\uparrow,p\!\downarrow,n\!\uparrow,n\!\downarrow)$ the Lagrangian $\Lag$
%contains all possible terms compatible with QCD symmetry ordered by the number
%of derivatives and nucleon fields. 
The resulting theory
coincides with  
%the pionless EFT 
$\nopi$EFT for natural nucleons 
\cite{paulo},
%\cite{nopi} up to a difference in the constituents mass 
%and the coupling constants.
except for different values of parameters and scales.
While in nature $\sqrt{m_N M_{nuc}}\sim 100$ MeV 
and $M\sim M_\pi$, 
for $m_{\pi}\sim 800\;\rm{MeV}$
the numbers in Table~\ref{tbl:EnergyScales} suggest instead  
$\sqrt{m_N M_{nuc}}\sim 200$ MeV 
and $M\sim \sqrt{M_{QCD}M_{\Delta}} 
%\sim \sqrt{m_N(m_\Delta-m_N)} 
\simeq 500$ MeV
({\it cf.} Ref. \cite{savagedeltas}).

%===========================================================
% The hamiltonian
%===========================================================

%{\it The Hamiltonian} -- 
%The EFT Lagrangian 
%$\Lag$
%contains all possible terms compatible with QCD symmetries
%and ordered by the number of derivatives and nucleon fields:
%The effective Lagrangian reads 
%\bea\label{LagLO} 
%{\cal L} &=& N^{\dagger}\left(i\partial_0+\frac{\vec\nabla^2}{2m_N}\right)N
%\nn[2pt] & &
%- c_1 (N^{\dagger}N)( N^{\dagger}N)
%- c_2 (N^{\dagger}\sgmvec N) \cdot (N^{\dagger}\sgmvec N)
%\nn[2pt] & &
%- d_1 (N^{\dagger}\vec{\tau} N) \cdot (N^{\dagger}\vec{\tau} N)N^{\dagger}N
%+\ldots
%\eea
%here ``$\ldots"$ stands for terms containing 
%different spin/isospin combinations, derivatives, and/or larger number
%of fields. The expansion coefficients $\{c_{1,2}(M,m_{\pi}),d_1(M,m_{\pi})\}$, 
%commonly called low-energy constants (LECs),
%are unknown parameters that depend on $M$ and $m_{\pi}$. They 
%encompass information about momenta higher than the breaking scale of the EFT, 
%and should be fitted through comparison between the EFT and the
%available data.   

{\it The Hamiltonian} -- 
The EFT Lagrangian contains all possible terms compatible with QCD symmetries
and ordered by the number of derivatives and nucleon fields.
%The effective Hamiltonian resulting from Eq. (\ref{LagLO}) 
The corresponding Hamiltonian 
is naturally formulated 
in momentum space, where the potential takes the form of a
momentum expansion that must be regulated in high momentum.
Dependence only on transferred momenta leads to
local interactions, while more general momentum
dependence yields non-local interactions as well.

%The two-body part of the interaction can be written in terms of the incoming
%momentum $\pvec$
%and outgoing momentum $\pvec'$ or, equivalently, in terms of the
%momentum transfer $\bs{q}=\pvec'-\pvec$ and $\bs{k}=(\pvec'+\pvec)/2$. 
%When transforming the potential into configuration space 
%the $\bs{k}$ dependence
%leads to 
%yields non-local interactions, whereas 
%the $\bs{q}$ dependence leads to 
%non-locality only for a $\bs{k}$-dependent regulator.

Due to the Pauli principle we need consider only antisymmetric multinucleon
states. Restricting the Lagrangian to this subspace we are free to choose
a subset of the terms in the EFT Lagrangian without loss of generality.
%in order to produce the most general dynamics.
%Eq. (\ref{LagLO}) as a basis for $\Lag$. 
%Here we choose as LO the interactions shown explicitly in Eq. (\ref{LagLO}).
As in Ref. \cite{Gezerlis13}, 
here we aim at formulating a local EFT nuclear
potential that will allow us to study the many-body problem
utilizing techniques, such as 
AFDMC \cite{AFDMC1,AFDMC2}, that are restricted to local interactions. 
In order not to introduce non-local terms in the regularization,
we assume a regulator function of Gaussian form,
$f_{\Lambda}(q)= \exp({-\bs{q}^2/\Lambda^2})$
in terms of the momentum transfer $\bs{q}$
%$\bs{q}=\pvec'-\pvec$ 
and
a regulator parameter (or cutoff) $\Lambda$.
%In coordinate space,
%\be \label{cutoff_r}
%F_{\Lambda}(r)=\int \frac{d\bs{q}}{(2\pi)^3} f_{\Lambda}(q)
%e^{i\bs{q}\cdot\bs{r}}
%= \left(\frac{\Lambda}{\sqrt{4\pi}}\right)^3e^{-\Lambda^2 r^2/4} \;.
%\ee

For this regulator the coordinate-space Hamiltonian takes the form 
\bea
 H &=& -\sum_i \frac{\nabla^2_i}{2 m_N} 
   + \sum_{i<j} \left(C_1 + C_2\,\sgmvec_i\cdot\sgmvec_j \right) 
               e^{-\Lambda^2r_{ij}^2/4}   \cr
   &&+ \sum_{i<j<k}\sum_{cyc} D_1 \left(\tauvec_i\cdot\tauvec_j \right) 
               e^{-\Lambda^2(r_{ik}^2+r_{jk}^2)/4} 
+\ldots, \;\;
\label{Heff}
\eea
where 
%the notation 
$\sum_{cyc}$ stands for the cyclic permutation of
a particle triplet $(ijk)$,
and ``$\ldots"$ for terms containing 
more derivatives and/or more-body forces.
The expansion coefficients 
$C_{1,2}(m_{\pi},\Lambda), D_1(m_{\pi},\Lambda), \ldots$, 
commonly called low-energy constants (LECs),
are unknown parameters that encompass physics
at the scale $M$ and above, and thus change with $m_{\pi}$.
They depend on the arbitrary cutoff $\Lambda$
in such a way that low-energy observables are (nearly) cutoff-independent,
and they should be fitted through comparison between the EFT and the
available data.  

Na\"ive scaling arguments suggest that the LECs should scale as 
$1/M^{1+d+\frac{3}{2}(n-4)}$, where $d$ is the number of derivatives and
$n$ is the number of nucleon fields \cite{Kaplan95}. 
However, the existence of shallow $S$-wave two-body bound states at 
$\sqrt{m_N M_{nuc}}\ll M$ requires enhancements in operators
that connect $S$ waves.
The LO two-body operators are those without derivatives \cite{Bira99}.
While a surprising enhancement in the non-derivative
three-body interaction promotes it to LO \cite{3BodyTerm}, 
the same is thought not to happen for four-body forces \cite{hammer4}.
More-body forces require derivatives and 
are expected to be further suppressed.
To match current lattice calculations we can neglect isospin violation.
For the first attempts at a description of real light nuclei 
with the leading interactions, see Ref. \cite{morepiless}.
The $m_\pi$ dependence of two- and three-nucleon observables
in $\nopi$EFT has been studied with input from chiral EFT
\cite{hammeretal}.

Some comments are in order about the EFT truncation.
%regarding the regulator. 
%{\it(i)} 
Expanding the regulator
%the regulator
around $q=0$, we see that 
it introduces terms of ${\cal O}(Q/\Lambda)^2$. 
%the order $(q/\Lambda)^2$. 
%{\it(ii)} 
Moreover, the regulator does not commute with the permutation operator,
%The commutation of $f_{\Lambda}(q)$  with the antisymmetrizer 
which gives rise to 
more general momentum dependence of the same order.
%$(q/\Lambda)^2$. 
%of the same order of magnitude.
%$\bs{k}$-dependent terms. 
These terms can be lumped with higher-order 
%terms 
interactions in the ``\ldots''
of Eq. \eqref{Heff}
without increasing the expected truncation error,
${\cal O}(Q/M)$, because we consider here $\Lambda \simge M$.
%These terms affect the potential at NLO level. 
%Since in the following we shall be working at LO,
%therefore it is 
%these terms are of the same order 
%as terms anyway neglected in the Lagrangian.  These terms are also
%of the order $(q/\Lambda)^2$ and therefore can also be neglected.
%Since we consider $\Lambda \simge M$, cutoff errors
%are expected to be no larger than 
%the intrinsic error introduced by truncation in Eq. \eqref{Heff}.
A conservative estimate of the truncation error
%${\cal O}(Q/M)$,
is about 40\%.

%Incorporating the prefactors $(\Lambda/\sqrt{4\pi})^3$ 
%of Eq. (\ref{cutoff_r}) into the LECs,
%the LO many-body coordinate-space Hamiltonian takes the form
%\bea
% H &=& -\sum_i \frac{\nabla^2_i}{2 m_N} 
%   + \sum_{i<j} \left(C_1 + C_2\,\sgmvec_i\cdot\sgmvec_j \right) 
%               e^{-\Lambda^2r_{ij}^2/4}   \cr
%   &&+ \sum_{i<j<k}\sum_{cyc} D_1 \left(\tauvec_i\cdot\tauvec_j \right) 
%               e^{-\Lambda^2(r_{ik}^2+r_{jk}^2)/4}\;,
%\eea
%where the notation $\sum_{cyc}$ stands for the cyclic permutation of
%particle triplet $(ijk)$.
%The LECs $C_{1,2}, D_1$ are functions of the arbitrary cutoff $\Lambda$
%in such a way that low-energy observables are (nearly) cutoff-independent.

%===========================================================
% Experimental input
%===========================================================
%{\it The lattice world} -- 
{\it Input data} -- 
The online publication of the NPLQCD data for the spectrum of the $A\leq 4$ 
nuclei last year \cite{NPLQCD13a}  
% was the driving force of
provided the motivation for the current work. 
The measured lattice binding energies of the deuteron, dineutron and 
triton, together
with that of the alpha particle, provided us with the three data points 
to which we fit our LO LECs,
plus one data point to validate it.
%, see Table \ref{tbl:Data}. 
In the
meanwhile new lattice results have appeared. 
Unquenched calculations of light nuclear
binding energies at $m_{\pi}=510\;\rm{MeV}$ were reported
\cite{Yamazaki12},
%, see Table \ref{tbl:Data}, 
and also the 
%NN 
two-nucleon (NN)
scattering lengths and effective ranges at $m_{\pi}=805\;\rm{MeV}$ 
\cite{NPLQCD13b}.
We assume here that the interaction has range $\sim 1/m_\pi$
and comparable effective ranges, but much larger scattering lengths.
Since the reported effective ranges are smaller than the scattering lengths,
our expansion should converge,
albeit at a slow rate.
Note, however, that
the data from Ref. \cite{NPLQCD13b} indicates an
almost degenerate double bound state pole in the NN $T$ matrix, 
which is thought to be 
%impossible to describe 
incompatible with a short-range non-relativistic potential
\cite{russians}.
Worse still, Ref. \cite{lqcd_nn_force} finds no 
%two-nucleon 
NN bound states
in a large range of pion masses that includes the values in 
Table \ref{tbl:Data}.
Until the dust settles, 
we concentrate on the 
%four 
LQCD data 
%points
in Table \ref{tbl:Data} 
for $m_{\pi}=805\;\rm{MeV}$, as a
first check of our proposed approach.

%===========================================================
% Computational tools
%===========================================================
{\it Calibration and predictions} --
For the calibration of the 
%two-body 
NN LECs we turn to the spin-isospin ($S$, $T$) basis
and define the channel constants $C_{S,T}\equiv C_{1}+[2S(S+1)-3]C_{2}$. 
We solve the 
two-body Schr\"odinger equation
using the Numerov method, and fit $C_{S,T}$ to the deuteron 
($S=1,\;T=0$) and
dineutron ($S=0,\;T=1$) binding energies.
%given in Table \ref{tbl:Data}.
To calibrate the 
%three-body contact coefficient 
LEC $D_1$ using the $^3$H binding energy $B_3$,
we 
%need to 
solve the three-body
Schr\"odinger equation with
%To this end we utilize 
the EIHH method, where
%In the former \cite{EIHH}, 
we expand the wavefunction into a set of
antisymmetrized hyperspherical-harmonics spin-isospin states.
Convergence is controlled by the hyper-angular quantum number
$K_{max}$, results being obtained by extrapolation to the limit
$K_{max}\to \infty$ \cite{EIHH}.
The corresponding error in our results
is estimated
to be smaller (for the lighter systems, much smaller) than
the EFT truncation error.

The LECs fitted to the central values
of the lattice results are presented in Table \ref{tbl:LECs}.
%Note that for the lowest cutoff we might expect relatively
%large cutoff effects, since it is not 
%large compared to 
%considerably smaller than 
%$M$. 
The cutoff dependence of the
%two-body 
NN LECs $C_{S,T}$ is
qualitatively similar to other regulators \cite{Bira99}.
We see no limit-cycle behavior \cite{3BodyTerm} in
%the three-body force parameter%
$D_1$,
possibly because our cutoff values are not large enough to probe 
the second branch of the periodic function.
%for the regulator function (\ref{cutoff_r}).

\begin{table}[bt]
\begin{center}
\caption{The 
%leading order 
LO LECs [GeV] for lattice nuclei at 
$m_{\pi}=805\;\rm{MeV}$,
as a function of the momentum cutoff $\Lambda$ [fm$^{-1}$].}
\label{tbl:LECs}
\begin{tabular}
{c@{\hspace{5mm}} c@{\hspace{5mm}} c@{\hspace{5mm}} c}\hline\hline
$\Lambda$ &  $C_{1,0}$  &  $C_{0,1}$ & $D_1$ \\
\hline
    2  &      $-$0.1480 &  $-$0.1382 &  $-$0.07515  \\
    4  &      $-$0.4046 &  $-$0.3885 &  $-$0.3902   \\
    6  &      $-$0.7892 &  $-$0.7668 &  $-$1.147   \\
    8  &      $-$1.302  &  $-$1.273  &  $-$2.648   \\
\hline\hline
\end{tabular}
\end{center}
\end{table}

A simple check of $\nopi$EFT
%pionless EFT 
at LO, which is equivalent to the large-scattering-length 
approximation to the two-body problem, is that for large cutoffs the 
%deuteron 
$S=1,\;T=0$ scattering length 
%$a_{31}$ 
is related  
to 
%its 
deuteron binding energy $B_{31}$ by 
$a_{31}\approx 1/\sqrt{m_N B_{31}}$ \cite{Bira99}. 
This relation suggests that $a_{31}$ should
approach $1.12\,\text{fm}$ for the lattice deuteron. 
For our Gaussian cutoff 
%function \eqref{cutoff_r},
%described above, 
we find $a_{31}=(1.2\pm 0.5)\,\text{fm}$, where we use 
the wide range of cutoff variation $2-14\,\text{fm}^{-1}$
to estimate the EFT error.
%\footnote{The number in parentheses denotes uncertainty due 
%to cutoff dependence.}
%$1.20(1)\,\text{fm}$. 
%The small deviation
%from the analytical result is due to the 
%cutoff dependence of the cutoff function, 
%as demonstrated by a use of 
With a sharper cutoff function 
$f_{\Lambda}(q)\to \exp({-\bs{q}^4/\Lambda^4})$ there
is quicker convergence to the expected number,
$a_{31}=(1.1\pm 0.1)\,\text{fm}$ in the same cutoff range.
%for which we get $a_{31}=1.13(1)\,\text{fm}$.

With LECs fixed, we now have a complete LO potential that 
%we
can be used to predict other properties of lattice nuclei.
As a first step in this direction we shall estimate the binding energies
$B_A$ for $A=4,5,6$.
%lattice nuclei. 
To solve the Schr\"odinger equation for these
systems we use, in addition to EIHH, also the AFDMC method.
%In the former \cite{EIHH}, we expand the wavefunction into a set of
%antisymmetrized hyperspherical-harmonics spin-isospin states.
%Convergence is controled by the hyper-angular quantum number
%$K_{max}$, results being obtained by extapolation to the limit
%$K_{max}\to \infty$.
In the latter technique \cite{AFDMC1,AFDMC2}, the ground-state energies are projected 
from an arbitrary initial state by means of a stochastic imaginary-time 
propagation. The numerical complexity related to the presence of 
operatorial terms in the interaction is reduced by using the 
Hubbard-Stratonovich transformation, at the price of introducing 
auxiliary fields as additional degrees of freedom.  

\begin{figure}[tb]\begin{center}
\includegraphics[width=8.6 cm]{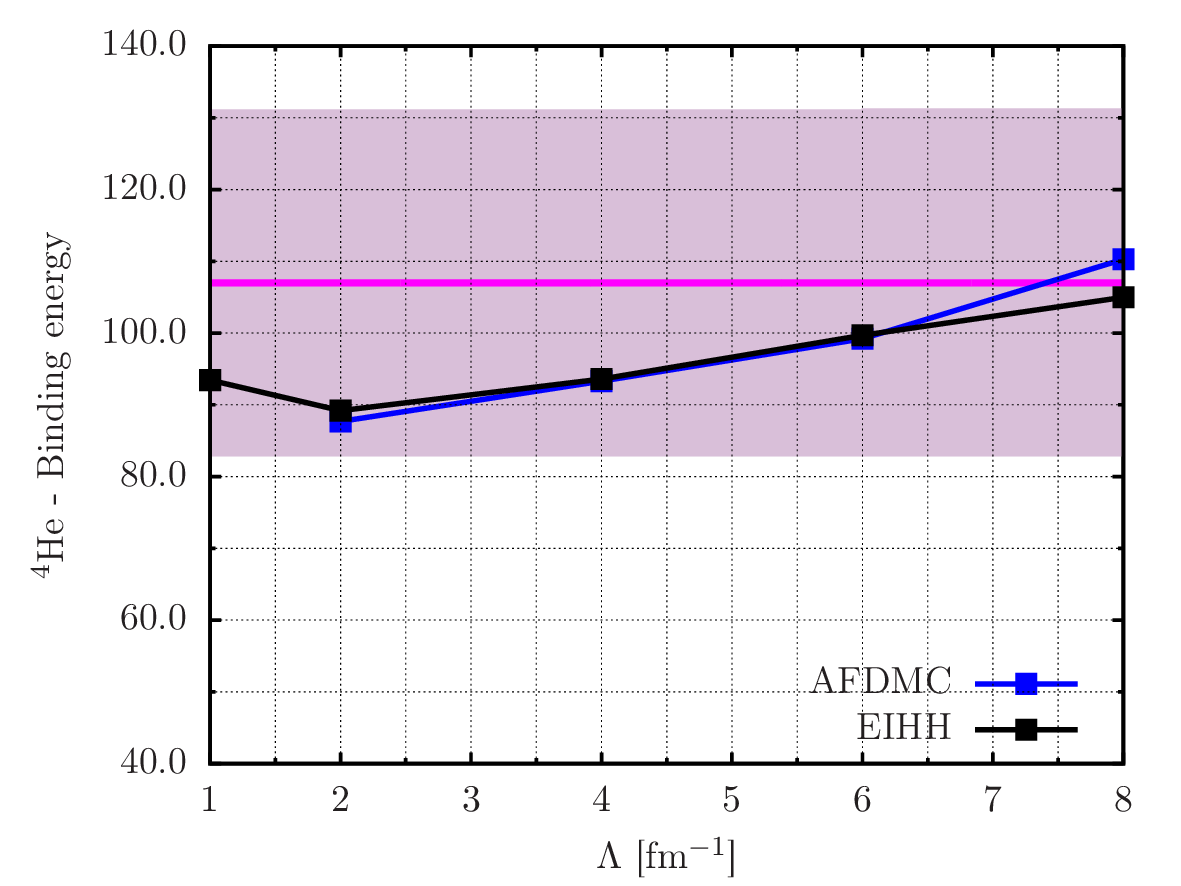}
\caption{\label{fig:4HeLmb} (Color online) $^4$He binding energy 
%at LO, 
$B_4$ [MeV] as function of the momentum cutoff $\Lambda$ [fm$^{-1}$]. 
The 
%black 
(magenta) horizontal line and the (pink) band give the 
%measured 
LQCD central value 
%result 
and error. The 
%blue 
(black and 
%purple
blue) solid lines are (respectively the EIHH and AFMDC)
%is the hyperspherical harmonics calculation and the purpule is the Monte Carlo
LO EFT results.}
\end{center}
\end{figure}

%In EFTs the LECs, and therefore also the observables at a given order,
%depend on the cutoff momentum $\Lambda$. 
In properly renormalized EFT, observables should be
asymptotically invariant to the value of $\Lambda$. 
In practice, this goal is hard
to achieve because in heavier systems our cutoff range is limited 
and variation of calculated properties tends to be smaller
than the expected EFT truncation error.
%as higher-order effects enter into the calculation.
%and we 
Therefore we
aim for a more modest goal, that observables depend only weakly on
the cutoff.  
In Fig. \ref{fig:4HeLmb} we present the calculated binding energy 
of $^4$He, $B_4$, as
a function of 
%the cutoff momentum 
$\Lambda$. 
%Comparing the calculated values
%with the measurement we see that 
Over a wide cutoff range
%of cutoff values, $2\; \rm{fm}^{-1}\leq \Lambda \leq 8\; \rm{fm}^{-1}$, 
%$2\leq \Lambda \; \rm{fm} \leq 8$, 
the EFT prediction reproduces the
LQCD result within the measurement error, evidence that
the EFT in LO captures the essence of the strong-interaction dynamics.
%From the figure 
%It is also evident that 
Our results are not 
cutoff invariant,
however the cutoff dependence is rather moderate:
%The binding energy 
$B_4$ changes by 
20$\%$ when $\Lambda$ grows by a factor of 4. 
%The LECs presented in Table \ref{tbl:LECs} were fitted to the central values
%of the lattice results, Table \ref{tbl:Data}. Considering the
%error associated with the lattice measurements we should bear in mind
%that an appropriate
%error estimate needs to be assigned to each of the LECs, and consequently also
%to the EFT predictions. 
%To demonstrate this point we present 

\begin{figure}[tb]\begin{center}
\includegraphics[width=8.6 cm]{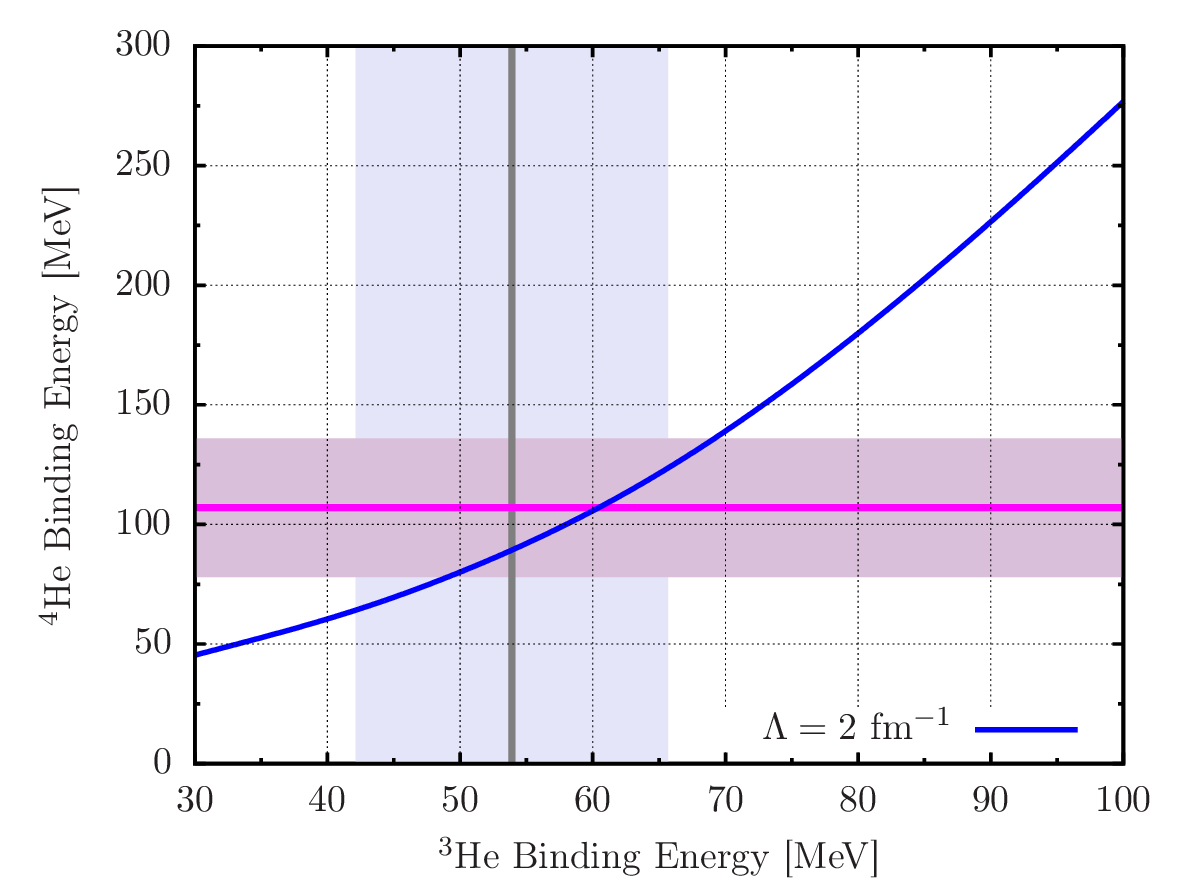}
\caption{\label{fig:B3B4} (Color online) 
Correlation between the $^4$He and $^3$He binding energies,
$B_4$ [MeV] and $B_3$ [MeV]. 
Horizontal and vertical lines and bands represent LQCD results.
The (blue) solid line is the Tjon line in LO EFT 
from EIHH at $\Lambda=2$ fm$^{-1}$.}
\end{center}
\end{figure}

In Fig. \ref{fig:B3B4} we present the correlation
between $^3$He and $^4$He binding
energies.
%as a function of 
%and the $^3$He binding energy, $B_3$. 
%From the figure 
When we allow $D_1$ to vary at fixed $C_{S,T}$, 
LO EFT gives a line \cite{hammer4},
which corresponds to the phenomenological Tjon line \cite{Tjon}.
One can
see how the measurement error in $B_3$ is propagated into an error in the
predicted $B_4$ value. 
Using this figure alone one would conclude that
%The resulting 
the EFT error estimate for $B_4$ is about
$\pm 27$ MeV, 
%This number is 
very close to the measurement error.
%and 
Our estimate of a 40\% error in the LO EFT is likely {\it very}
conservative.
Our reproduction of the LQCD central value and error estimate
for $B_4$ indicates consistency 
%between 
of the LQCD values \cite{NPLQCD13a}
%error estimates 
for the $A=2,3,4$
%and $A=4$ 
systems.

The power of the EFT 
%derived 
formulated above is the relative ease 
%through 
with which it can
be extended to 
%study 
different few- and many-body systems. 
Using 
%the cutoff 
$\Lambda=2\;\rm{fm}^{-1}$ we have looked
for excited states in $A=2,3,4$ systems, but much to our surprise
found none. Similarly, we have found no evidence for $^3$n droplets,
for which our
ground-state binding energy coincides with the two-body threshold.
Results for the $A=5,6$ ground states at $\Lambda=2\;\rm{fm}^{-1}$
are shown in Table \ref{tbl:Data},
with errors estimated from the EFT truncation.

%studied the existence of 
%excited 2-body states, $^3$n droplets, higher 
%excited states for in $A=3,4$ system and the
%ground states of the 5, and 6-body systems.
%Much to our surprise we didn't manage to identify any bound excited state in
%the 2, 3, and 4-body systems. Same goes with the $^3$n system, where the
%ground-state binding energy coincides with the 2-body threshold.

For 
%the 5-body system 
$^5$He we found a bound state 
%for $\Lambda=2\;\rm{fm}^{-1}$ 
with binding energy $B_5=98.2$ MeV,
which 
%This value 
reflects a 9 MeV binding relative to $^4$He at the same cutoff.
%the alpha particle. 
%This is in contrast with the
%natural case where $^5$He is unbound. 
However,
examining the evolution of $B_5$
%the $^5$He binding energy 
with the cutoff
%$\Lambda$ 
we found that for $\Lambda=4\;\rm{fm}^{-1}$ 
the five-body ground-state energy coincides with the four-body threshold.
This suggests that the $A=5$ nuclear gap found in nature persists
for larger quark masses.
%at this pion mass $^5$He is unbound as in nature.
%Therefore
%at this point we are inconclusive regarding the stability of $^5$He.    

%Considering now the 6-body system, 
We have also calculated the $^6$Li ground state
for $\Lambda=2\;\rm{fm}^{-1}$,
%using the EIHH method. 
%The convergence of the
%EIHH method is controled by the hyper-angular quantum number
%$K_{max}$. Extrapolating our results to the limit
%$K_{max}\longrightarrow\infty$
%we obtain 
obtaining $B_6\approx 122\;\rm{MeV}$.
In this case the error in $K_{max}$ extrapolation 
is about 3 MeV, which
is somewhat larger than for lighter systems but still
small compared with input and truncation errors.
%a binding energy value $B_6\approx 126\pm 5 \;\rm{MeV}$.
Thus $B_6/A \approx 20 \;\rm{MeV}$, similar to 
lattice $^4$He.
%for which $B_4/A\approx 22 \;\rm{MeV}$. 
We conjecture that nuclear saturation
survives the increase in pion mass, 
but this conclusion  
%Whether this indicates nuclear saturation in the lattice
%still 
remains to be 
%checked 
confirmed by larger calculations. 
{\it Conclusion} --
One of the main challenges of current research in nuclear physics is to 
provide a unified look at the nuclear regime, from QCD to heavy nuclei. 
Using results from recent lattice QCD simulations 
of few-nucleon systems, we took important steps in this direction
by demonstrating the consistency of $\nopi$EFT and LQCD
for $m_{\pi}\approx 800 \;\rm{MeV}$.
%The validity of $\nopi$EFT for lattice nuclei, similar to the natural case, 
%by fitting the nuclear Hamiltonian to LQCD data, and finding that 
%it reproduces the $^4$He binding energy for $m_{\pi}=800\rm{MeV}$ nucleons. 
%This also provides an additional check of the consistency 
%of the various LQCD calculations. 
Our results suggest that some of the defining properties of nuclei 
might be relatively insensitive to the value of the pion mass.
More LQCD data are needed in order to go beyond LO EFT
%in order to assess 
and assess the systematic uncertainty of the EFT approach,
%itself.
%Going beyond LO might be needed also for the
%prediction of other observables, such as radii.
while more extensive calculations with the EFT should
settle the issue of the importance of quark masses to nuclear properties,
with implications for the analysis of fundamental constant variability
\cite{funcons}.

%We believe that the combination of LQCD and EFT approaches for light nuclei 
%will prove as a useful tool, for example to extract more information from 
%lattice simulations. Moreover, the solution of the
%corresponding Schroedinger equation in a finite volume, can be used 
%to analyse lattice simulations and their extrapolation to larger volumes, 
%which today can be done only by increasing the lattice size. 

\begin{acknowledgments}
We thank the National Institute for Nuclear Theory at the
University of Washington for the hospitality when this work
was conceived (INT-PUB-13-043),
and Silas Beane and Martin Savage for useful discussions.
This research was supported by 
the Israel Science Foundation under grant number 954/09 (NB), 
a BMBF ARCHES grant (DG),
the US DOE under grant DE-FG02-04ER41338 (UvK),
and 
the ``Attractivit\'e 2013'' program of the Universit\'e Paris-Sud (UvK). 
FP and LC are members of LISC, the Interdisciplinary Laboratory for 
Computational Science, a joint venture of the University of Trento 
and FBK (Bruno Kessler Foundation).
\end{acknowledgments}

%===========================================================
% Bibliography
%===========================================================

\end{document}